%% file: main.tex
\documentclass[conf]{new-aiaa} 
\usepackage[utf8]{inputenc}

\usepackage{graphicx}
\usepackage{amsmath}
\usepackage{comment}
\usepackage{siunitx}
\usepackage{tikz}
\usetikzlibrary{arrows, positioning}
\input{PackagesCommands}
\newcommand{\hinf}{$\SH_2/\SH_\infty \ $}

\usepackage[colorinlistoftodos,textwidth=1 in,textsize=footnotesize]{todonotes}
\title{Low-Order \hinf Controller Design for \\ Aeroelastic Vibration Suppression}

\author{
Mohammad Mirtaba%
\footnote{Graduate Research Assistant, Department of Mechanical Engineering, University of Maryland, Baltimore County, 1000 Hilltop Circle, Baltimore, MD 21250. \texttt{mmirtab@umbc.edu}}
,
Juan Augusto Paredes Salazar%
\footnote{Postdoctoral Research Fellow, Department of Mechanical Engineering, University of Maryland, Baltimore County, 1000 Hilltop Circle, Baltimore, MD 21250. \texttt{japarede@umbc.edu}}
, 
Daning Huang%
\footnote{Assistant Professor, Department of Aerospace Engineering, Pennsylvania State University, 556 White Course Drive, University Park, PA 16802. \texttt{daning@psu.edu}}
, and Ankit Goel
\footnote{Assistant Professor, Department of Mechanical Engineering, University of Maryland, Baltimore County, 1000 Hilltop Circle, Baltimore, MD 21250. \texttt{ankgoel@umbc.edu}}
}

\begin{document}

\maketitle





\begin{abstract}
    This paper presents an \hinf minimization-based output-feedback controller for active aeroelastic vibration suppression in a cantilevered beam. 
    %
    First, a nonlinear structural model incorporating moderate deflection and aerodynamic loading is 
    derived and discretized using the finite element method (FEM).
    %
    %
    Then, a low-order linear model is identified from random gaussian input response data from the FEM model to synthesize an output-feedback controller using the \hinf framework.
    A frequency-weighted dynamic filter is introduced to emphasize disturbance frequencies of interest, enabling the controller to target dominant vibration modes. 
    %
    %
    Simulation results demonstrate the effectiveness of the proposed technique for vibration suppression and study its robustness to system parameter variations, including actuator placement.
    %
\end{abstract}

\section{Introduction}

Flexible aerospace structures such as wings, panels, and control surfaces are susceptible to vibrations induced by aerodynamic forces.
These vibrations can lead to degraded performance, structural fatigue, or instability—particularly in regimes where aeroelastic effects such as flutter become significant \cite{chai2021aeroelastic}.
To ensure structural integrity and operational reliability, active vibration suppression strategies must be used \cite{hu2005, hu2009, he2015, tsushima2018, prakash2016, bloemers2024}.
Several control schemes based on the classical LQR and robust control have been explored for this problem \cite{zhang2008,schulz2013,zhang2024,souza2019,fan2019,wang2022}.
%
Among the most powerful frameworks for robust control in uncertain and dynamic environments is the \hinf optimal control approach, which balances disturbance attenuation with stabilization performance. 

Motivated by the flutter problem, this paper focuses on the design and implementation of an \hinf output-feedback controller for suppressing vibrations in a cantilevered beam subjected to aerodynamic excitation.
Furthermore, a frequency-weighted design is adopted to target specific disturbance frequencies, ensuring that the controller focuses on the dominant vibratory modes.
%
%
%
To model the cantilevered beam, a structural model is developed, which includes geometric nonlinearities due to moderate deflections and incorporates aerodynamic loading derived from piston theory \cite{Dowell1974}.
The resulting nonlinear partial differential equations are discretized using the finite element method (FEM) to obtain a high-fidelity structural model.
%
%
A low-order linear approximation is identified from time-domain random gaussian input response data from the FEM model to obtain a linear model, which is used to synthesize an output-feedback controller using the \hinf framework.
The \hinf controller design framework is applied to vibration attenuation in the FEM model through numerical simulations.
Two cases are evaluated, in which the controller is used to attenuate the cantilever tip displacement vibrations caused by an external harmonic disturbance and by aerodynamic loading.
Note that the external harmonic disturbance case is used to preliminarily evaluate the proposed technique, and the aerodynamic loading is applied to induce flutter. 
%
%

The paper is organized as follows.
Section \ref{sec:StructModel} describes the structural modeling and discretization process.
Section \ref{sec:control} outlines the \hinf control formulation and design methodology. 
Section \ref{sec:simulations} presents numerical simulations validating the performance of the controller, as discussed before.
Conclusions and future work are discussed in Section \ref{sec:conclusions}.

\section{Structural Dynamics Model} \label{sec:StructModel}

The section briefly reviews the structural dynamics model of a cantilevered beam and presents the finite-element model used to simulate the beam in this work. 
Subsection \ref{subsec:gov_eq} presents the governing equations corresponding to a nonlinear beam model with moderate deflection.
Subsection \ref{subsec:fem} presents the beam equation solution using FEM, which results in the model used for numerical simulations.

\subsection{Governing Equation}\label{subsec:gov_eq}

In this study, a nonlinear beam model with moderate deflection is considered for a beam of length $L.$
The governing equation, as shown in \cite{Friedmann2023}, is
\begin{equation}\label{eqn_beam_pde}
    m\ddot{w} + \frac{\partial }{\partial x} \left(N \frac{\partial  w}{\partial x}\right) + D \frac{\partial^4 w}{\partial x^4} = p \left(x,t,w,\frac{\partial w}{\partial x},\dot{w}\right),
\end{equation}
where $x \in [0, L]$ denotes a position along the beam, $w$ is the transverse displacement, $m$ and $D$ are the mass per unit length and the bending stiffness, respectively, and
%
%
$$
N[w] \isdef \int_0^L EI \left(\frac{\partial w}{\partial x}\right)^2 dx
$$
is the in-plane force that accounts for the nonlinear effect due to moderate deflection, where $E$ is the Young's modulus corresponding to the beam material and $I$ is the inertial of the beam along its longitudinal axis. 
The external load $p$ can be further decomposed as
\begin{align}
    p\left( x,t,w,\frac{\partial w}{\partial x},\dot w \right)
        =
            p_\rme(x,t) + 
            p_\rmc(x,t) + 
            p_\rmd(\dot w) + 
            p_\rma\left( x,\frac{\partial w}{\partial x}, \dot w \right),
\end{align}
where
$p_\rme(x,t) $ is an external excitation of the form
\begin{align}
    p_\rme(x,t)
        =
            \begin{cases}
                p_\rme(t), & x\in (\ell_1, \ell_2) \subset [0, L], \\
                0 , & \rm otherwise,
            \end{cases}
\end{align}
$p_\rmc(x,t)$ is the control force given by
\begin{align}
    p_\rmc(x,t)
        =
            \begin{cases}
                p_\rmc(t), & x = x_\rmc, \\
                0 , & \rm otherwise,
            \end{cases}
\end{align}
where $x_\rmc \in [0, L]$ is the point of application of the control, 
%
%
$p_\rmd(\dot w) = -\zeta \dot w$ is a damping force, where $\zeta$ is a coefficient of damping, 
and 
$p_\rma(x,w_x, \dot w)$ is the aerodynamic load, based on the piston theory for high-speed flow \cite{Dowell1974}, which can be modeled as
%
%
\begin{align}
    p_\rma \left(x, \frac{\partial w}{\partial x} ,\dot{w}\right) = p_\infty \left( 1 + \frac{\gamma-1}{2}M_n \right)^{\frac{2\gamma}{\gamma-1}} - p_\infty,
\end{align}
where 
$p_\infty$ is the free-stream static pressure,
$\gamma$ is the specific heat ratio of air ,
%
and $M_n \isdef M_\infty \frac{\partial w}{\partial x} + \frac{\dot{w}}{a_\infty},$
where $M_\infty$ is the free-stream Mach number,
and $a_\infty$ is the free-stream speed of sound. 
Note that the external load $p$ may include the full or partial combination of loads above. 
%
%
%
%
%
%
Since this work considers a cantilevered beam, the boundary conditions are given by
\begin{equation}\label{eqn_beam_bc1}
    w(0)=\frac{\partial w}{\partial x}(0)= \frac{\partial^2 w}{\partial x^2}(0)=\frac{\partial^3 w}{\partial x^3}(0)=0.
\end{equation}


\subsection{Finite Element Model} \label{subsec:fem}

The beam equation is solved by the finite element method (FEM) using standard third-order Hermite polynomials.  Formally, the structural deformation is discretized as
\begin{equation}
w(x,t) = \sum_{i=1}^N \phi_i(x)u_i(t) \equiv \phi^\top u,
\end{equation}
where $\phi_i(x)$ are the Hermite shape functions defined on each of the elements, and $u_i(t)$ physically represents the displacement and rotation at the nodes of the elements. 
Next, \eqref{eqn_beam_pde} can be written as 
\begin{equation}\label{eqn_beam_fe}
    M \ddot{u} + K(u) u = f(u,t),
\end{equation}
where $M\isdef \int_0^L m \phi\phi^\top dx$ is the the mass matrix,
$K(u)=K_\rmL+K_\rmN(u)+K_\rmb$ is the stiffness matrix, where the linear and nonlinear terms are, respectively,
$$
K_\rmL=\int_0^L D \left(\frac{\partial^2 \phi}{\partial x^2}\right) \left(\frac{\partial^2 \phi}{\partial x^2}\right)^\top \rmd x,\quad K_\rmN(u)=\int_0^L \left(\frac{\partial \phi}{\partial x}\right) \left(\frac{\partial \phi}{\partial x}\right)^\top N[\phi^\top u] \rmd x,
$$
and the term $K_\rmb$ enforces the boundary conditions by penalty method, and is determined by Eq.~\eqref{eqn_beam_bc1}.
Lastly, the forcing vector is $f(u,t)=\int_0^L \phi p dx$; for the convenience of later discussion, it is written as
$$
f(u,t) = f_1(u,t) + f_\rme p_\rme(t) + f_\rmc p_\rmc(t)
$$
where $f_1$ includes the damping and aerodynamic forces, and $f_\rme$ and $f_\rmc$ are due to excitation and control, respectively.
Finally, the measurements are computed via the finite element interpolation, that is,  at location $x^*$ and time $t^*$, the displacement is
%
%
\begin{align}
    w (x^*, t^*) = \phi(x^*)^\top u(t^*).
\end{align}
%





\section{ \hinf Control} \label{sec:control}
This section provides a brief review of the standard control problem and presents the \hinf control objective and implementation details.
Subsection \ref{subsec:control_problem} reviews the standard control problem and presents an augmented closed-loop transfer function used for \hinf design.
Subsection \ref{subsec:hinf_control_objective} presents the \hinf control objective and controller design details.
Subsection \ref{subsec:sys_id} introduces a system identification procedure to obtain a system model for the design of the \hinf controller.
Subsection \ref{subsec:implementation_details} provides the \hinf controller design and implementation details for interfacing the controller with the FEM simulation.

\subsection{Standard Control Problem} \label{subsec:control_problem}

Consider the system 
\begin{align}
\dot{x}(t) &=  A x(t) + B u(t) + D_1 w(t), \label{plant}\\
y(t) &=  C x(t) + D u(t) + D_2 w(t), \label{outputy}\\
z(t) &= E_1 x(t) + E_2 u(t),\label{z2}
\end{align}
where 
$x \in \BBR^{l_x}$ is the state, 
$u \in \BBR^{l_u}$ is the control signal,
$w \in \BBR^{l_w}$ is the exogenous signal, 
$y \in \BBR^{l_y}$ is the measurement, and 
$z \in \BBR^{l_z}$ is the performance variable. 
Consider the $n$th-order dynamic output-feedback controller
\begin{align}
    \dot{x}_\rmc(t)&=  A_\rmc x_\rmc(t) + B_\rmc y(t), \label{controllerplant} \\
    u(t)&=  C_\rmc x_\rmc(t), \label{controlleroutput}
\end{align}
where $x_\rmc \in \BBR^{l_c}$ is the controller state.
The closed-loop system \eqref{plant}--\eqref{controlleroutput} is given by
\begin{align}
    \dot{\tilde{x}}(t) &= \tilde{A} \tilde{x}(t) + \tilde{D} w(t),\label{closed_loop_system}
    \\
    z(t) &= \tilde{E} \tilde{x}(t).
\end{align}
where
\begin{gather}
\tilde{x}(t) \isdef \matl  x(t) \\ x_\rmc(t) \matr, 
\quad \tilde{A} \isdef \matl  A & B C_\rmc \\ B_\rmc C & A_\rmc + B_\rmc D C_\rmc \matr,
\quad \tilde{D} \isdef \matl  D_1 \\ B_\rmc D_2 \matr,\label{closed_loop_system.2}
\quad \tilde{E} \isdef \matl  E_1 & E_2 C_\rmc \matr.
\end{gather}

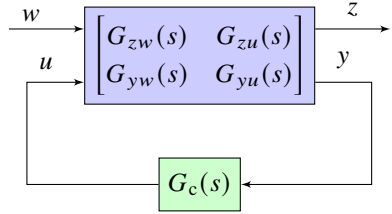
\begin{figure}[h!]
    \centering
    {
    \begin{tikzpicture}[auto, node distance=2cm,>=latex',text centered]

    \node [smallblock] (Plant) 
        {$
        \matl
                G_{zw}(s)  & G_{zu}(s) \\ 
                G_{yw}(s)  & G_{yu}(s) 
            \matr
        $};

        \node [smallblock, fill=green!20, below = 2 em of Plant] (Gc) 
        {$
        G_\rmc(s)
        $};

        \draw[->] (Plant.13) -- node[xshift = 0em, yshift = 0.15em]{$z$} +(1,0) ;
        \draw[->] (Plant.-13) -- node[xshift = 0em, yshift = 0.15em]{$y$} +(0.75,0) |- (Gc.0); 

        \draw[<-] (Plant.167) node[xshift = -2em, yshift = 0.5em]{$w$} -- +(-1,0) ;
        \draw[->] (Gc.180) -- +(-1.75,0) |- node[xshift = 0.75em, yshift = 0.15em]{$u$} (Plant.193) ; 
        
    \end{tikzpicture}
    }
    \caption{Standard control architecture with transfer function representation. }
    \label{fig:SCA}
\end{figure}

Note that 
\begin{align}
    G_{yw}(s) &= C (sI-A)\inv D_1 + D_2, \quad 
    G_{yu}(s) = C (sI-A)\inv B, \\
    G_{zw}(s) &= E_1 (sI-A)\inv D_1, \quad 
    G_{zu}(s) = E_1 (sI-A)\inv B + E_2, 
\end{align}
and thus
\begin{align}
    \matl
        Z(s) \\
        Y(s)
    \matr
    =
        \SG(s)
        \matl
            W(s) \\
            U(s)
        \matr,
\end{align}
where
\begin{align}
    \SG(s)
        \isdef
            \matl
                G_{zw}(s)  & G_{zu}(s) \\ 
                G_{yw}(s)  & G_{yu}(s) 
            \matr.
\end{align}
Figure \ref{fig:SCA} shows a block diagram of the closed-loop system \eqref{plant}--\eqref{controlleroutput} with $\SG(s)$ and $G_\rmc (s),$ which is the controller transfer function corresponding to \eqref{controllerplant}, \eqref{controlleroutput}.
Note that the closed-loop transfer function from $w$ to $z$ is given by
\begin{align}
    \tilde G_{z w}(s) \isdef \tilde E (s I_{l_x + l_c} - \tilde A)^{-1} \tilde D = G_{z w} (s) + G_{z u} (s) (I_{l_u} - G_\rmc(s) G_{y u} (s))^{-1} G_\rmc (s) G_{y w} (s). \label{eq:tildeGzw}
\end{align}

\subsubsection{Closed-loop transfer function augmentation with frequency-weighting dynamic filters} \label{subsubsec:tf_filters}

It is well known that the response of a linear system to a harmonic input is a harmonic signal of the same frequency, differing only in amplitude and phase.
%
%
Therefore, to minimize the effect of a specific disturbance frequency, frequency-weighting dynamic filters are used to emphasize the response of the system at the disturbance frequency so that the \hinf controller targets it. 
In particular, we filter the measurement and control signals as 
\begin{align}
    y_\rmf = W_y(s) y, \quad u_\rmf = W_u(s) u,
\end{align}
where $W_y, W_u$ are proper and stable transfer functions.
Then, it follows from \eqref{eq:tildeGzw} that the closed-loop transfer function from $w$ to $z$ augmented with the filter is given by
\begin{equation}
    \tilde G_{{z w}, {\rm aug}}(s) = G_{z w}(s) + G_{z u}(s) W_u (s) (I_{l_u} - G_\rmc(s)  W_y (s) G_{y u}(s) W_u (s) )^{-1} G_\rmc (s) W_y (s) G_{y w} (s). \label{eq:tildeGzw_aug}
\end{equation}

\subsection{\hinf Control Objective} \label{subsec:hinf_control_objective}

The objective of the { \hinf-constrained linear-quadratic-Gaussian control problem} is to determine the controller \eqref{controllerplant}, \eqref{controlleroutput} such that the following are satisfied
\begin{enumerate}
	
	\item $\tilde A$ is asymptotically stable.
	\item $\tilde G_{{z w}, {\rm aug}}$ satisfies the $\SH_\infty$ constraint given by
	\begin{align} \label{eq:G_infty}
	   \Vert \tilde G_{{z w}, {\rm aug}}\Vert_\infty \isdef \sup_\omega \sigma_{\rm max} (\tilde G_{{z w}, {\rm aug}} (\jmath \omega)) \leq \gamma_\infty ,
	\end{align}
	where $\sigma_{\rm max} (\tilde G_{{z w}, {\rm aug}} (\jmath \omega))$ is the maximum singular value of  $\tilde G_{{z w}, {\rm aug}} (\jmath \omega),$ and $\gamma_\infty >0$ is a given constant.
	\item Minimization of the $\SH_2$ cost given by
	\begin{align} \label{eq:G_2}
	   J(A_\rmc, B_\rmc, C_\rmc) \isdef \Vert \tilde G_{{z w}, {\rm aug}}\Vert_2^2 \isdef \frac{1}{2\pi} \int_{- \infty}^{\infty} \Vert \tilde G_{{z w}, {\rm aug}} (\jmath \omega) \Vert_\rmF^2 \ \rmd \omega,
	\end{align}
	where $\Vert \tilde G_{{z w}, {\rm aug}} (\jmath \omega) \Vert_\rmF$ is the Frobenius norm of $\tilde G_{{z w}, {\rm aug}} (\jmath \omega).$
\end{enumerate}

The controller construction is described in more detail in \cite{doyle1988state}.
%
%
The \hinf framework is an extension of the classical LQG framework with the additional constraint on the frequency response of the closed-loop transfer function, as is shown above.
This constraint allows the suppression of specific frequencies in the closed-loop.
However, note that, unlike the LQG controller whose existence is guaranteed if $(A,B)$ is stabilizable and $(A,C)$ is observable, the existence of an output feedback controller such that the user-defined $\SH_\infty$ constraint is satisfied is not guaranteed. 
In practice, the $\SH_\infty$ constraint is successively relaxed until it is satisfied. 
In this work, we design and obtain the output feedback controller by using the MATLAB function \href{https://www.mathworks.com/help/robust/ref/dynamicsystem.hinfsyn.html}{hinfsyn}, which tries to minimize $\gamma_\infty.$

\subsection{System Identification for Controller Design} \label{subsec:sys_id}

The design of the \hinf controller requires the knowledge of closed-loop transfer function from $w$ to $z,$ as mentioned above. 
However, this information is usually not available.
%
%
%
%
%
Hence, a low-order linear system is identified by collecting the input-output data of the beam excited with random gaussian signal.

We consider sampled measurement and input signals for system identification, and we assume that $y, u \in \BBR.$
Hence, we define the sampled measurement $y_k \isdef y(k T_\rms)$ and the sampled input $u_k \isdef u(k T_\rms),$ where $k \ge 0$ is sample step and $T_\rms > 0$ is the sampling period.
Next, to construct an $n$th-order linear model from input-output data, we consider the $n$th-order linear difference equation
\begin{align}
    y_{k+n}
        =
            a_{n-1} y_{k+n-1} + \ldots 
            a_{0} y_{k} + 
            b_{n-1} u_{k+n-1} \ldots 
            b_{0} u_{k}, \label{eq:lin_model}
\end{align}
where $a_0, \ldots, a_{n-1}, b_0, \ldots, b_{n-1} \in \BBR$ are the linear model coefficients.
Note that \eqref{eq:lin_model} can be written in the transfer function form as 
\begin{align}
    y_k
        =
            \frac
                {b_{n-1} \shiftq^{n-1} + \ldots b_0}
                {\shiftq^n - a_{n-1} \shiftq^{n-1} + \ldots a_0}
            u_k, 
\end{align}
where $\shiftq$ is the forward shift operator, 
and in the regressor form as
\begin{align}
    y_{k+n}
        =
            \phi_k \theta, \label{eq:IOModel}
\end{align}
where 
\begin{align}
    \phi_k
        \isdef 
            \matl 
                y_{k+n-1} &
                \cdots &
                y_{k} &
                u_{k+n-1} &
                \cdots &
                u_{k} 
            \matr
            \in \BBR^{1 \times 2n}, 
    \quad 
    \theta
        \isdef 
            \matl 
                a_{n-1} &
                \cdots &
                a_{0} &
                b_{n-1} &
                \cdots &
                b_{0} 
            \matr^\top
            \in \BBR^{2n}.
\end{align}
It follows from \eqref{eq:IOModel} that 
\begin{align}
    Y = \Phi \theta,
    \label{eq:AugmentedLRE}
\end{align}
where
\begin{align}
    Y
        \isdef
            \matl 
                y_{k + n} \\
                y_{k + n + 1} \\
                \vdots \\
                y_{k + n + N  -1} 
            \matr \in \BBR^N,
    \quad 
    \Phi
        \isdef
            \matl 
                \phi_{k} \\
                \phi_{k + 1} \\
                \vdots \\
                \phi_{k + N - 1} 
            \matr
            \in \BBR^{N \times 2n},
\end{align}
where $N \ge 1.$
Assuming that $\Phi$ is full-column rank, the least-squares solution of \eqref{eq:AugmentedLRE} is given by
\begin{align}
    \theta 
        =
            (\Phi^\rmT \Phi)\inv \Phi^\rmT Y. \label{eq:theta}
\end{align}
In the context of system identification, \eqref{eq:theta} shows the calculation to obtain the coefficients of a linear model from $n + N$ measurement and input samples.
To test the accuracy of the identified model, 
The accuracy of the identified model can be assessed by calculating the RMSE cost given by
\begin{align}
    J &\isdef \sqrt{ \frac{1}{N} \sum_{i=n}^{n + N - 1} \left( y_{k + i} - \phi_{k + i - n} \ \theta \right)^2 }.
\end{align}
%
%

\subsection{Controller Design and Implementation Details for Interface with Diffuser Model Simulation} \label{subsec:implementation_details}

In all numerical simulations, the measurement is the displacement at the tip of the cantilever beam, such that $y (t) =  w (L, t),$
and the input is the control force and, unless otherwise stated, it is applied at the tip as well, such that $u (t) = p_\rmc (t)$ and $x_\rmc = L.$
To improve the accuracy of the identified model, the input and output data is generated by exciting the cantilever beam FEM with random gaussian input in $u$ identify $G_{yu}.$

Since the objective is to minimize the vibrations in $y,$ we set $z = y.$
Furthermore, in practice, determining the impact of the disturbance $w$ on $y$ and $z$ is a complex task.
Hence, for practical purposes, we assume that $G_{zw} = G_{zu}$ and $G_{y_w} = G_{yu},$ which can be interpreted as assuming that the input and the disturbance are applied at the same location in the cantilever.
The validity of this last assumption is shown in the numerical simulation results in Section \ref{sec:simulations}.

Since the system identification procedure yields a sampled-data model, we proceed by designing a sampled-data \hinf controller that requires the derived $G_{yu}(\shiftq)$ and the sampled-data filters $W_y (\shiftq)$ and $W_u (\shiftq),$ which are designed in continuous-time and discretized using the \texttt{c2d} Matlab command.
These are used with the using the MATLAB function \href{https://www.mathworks.com/help/robust/ref/dynamicsystem.hinfsyn.html}{hinfsyn}, since it also allows the design of sampled-data controllers.

The \hinf controller is implemented as a sampled-data controller, with the zero-order-hold (ZOH) and the sampler, as digital-to-analog (D/A) and analog-to-digital (A/D) interfaces, respectively.
The controller samples the measurement signal and issues a control signal every $T_\rms$ s, which corresponds to an increase of the sampled time step $k.$  
Hence, at time step $k,$ the \hinf controller samples the displacement of the tip of the cantilever, such that
\begin{equation}
    y_k = w (L, k T_\rms),
\end{equation}
and modulates the control force, such that, for all $t \in [k T_\rms, (k+1) T_\rms),$
\begin{equation}
    p_\rmc(t) = u_k.
\end{equation}

\section{Numerical Simulation Results} \label{sec:simulations}
%
%
This section applies the \hinf minimization framework to design an output feedback controller to suppress vibrations in a cantilever beam. 
To simulate the cantilever beam with the properties described in Table \ref{fig:table_property}, the system is discretized into 20 elements, which results in 42 degrees of freedom.  
When a nonlinearity is involved, such as in $K_\rmN$ and $f$, 7th-order Gaussian quadrature is employed to numerically evaluate the corresponding integrals.
Once the discretized system ~\eqref{eqn_beam_fe} is obtained, the generalized-$\alpha$ method is employed for time integration; the solution is considered the full-order solution.

Two cases are evaluated.
In the first case, the \hinf controller is used to attenuate the cantilever tip displacement vibrations caused by an external harmonic disturbance.
In the second case, the controller is used to attenuate the cantilever tip displacement vibrations caused by aerodynamic loading.
Note that the external harmonic disturbance case is used to preliminarily evaluate the proposed technique, and the aerodynamic loading is applied to induce flutter. 
In all cases, the open-loop results (without control) are compared with the closed-loop results (with the \hinf controller).

\begin{table}[h!]
\centering
\caption{Cantilever Beam Material and Geometric Properties}
\begin{tabular}{l c c}
\hline
\textbf{Property} & \textbf{Symbol} & \textbf{Value} \\
\hline
Young's modulus      & $E$     & $\SI{1e9}{\pascal}$ \\
Poisson's ratio      & $\nu$   & $0.3$ \\
Density              & $\rho$  & $\SI{1}{\kilogram\per\meter\cubed}$ \\
Structural damping ratio    & $\zeta$ & 0.05 \\
Beam thickness (unit width) & $h$ & $\SI{0.002}{\meter}$ \\
Beam length          & $L$     & $\SI{1}{\meter}$ \\
\hline
\end{tabular}
\label{fig:table_property}
\end{table}


\subsection{Harmonic Disturbance}
A harmonic disturbance $w_k$ is applied through $p_\rme$ at $x \in (\ell_1, \ell_2) = (0.7, 0.8),$
where for all $k \geq 0,$
\begin{align}
    w_k = 10^{-3} \sin \left( {\dfrac{\pi}{3}k} \right),
\end{align}
and, for all $t \in [k T_\rms, (k + 1) T_\rms],$
\begin{equation}
    p_\rme(t) = w_k.
\end{equation}
In the simulations in this subsection, the sampling time is $T_\rms = 5 \times 10^{-3}$ s. 
The objective is to reduce the vibrations at the tip of the beam, that is, at $x = L = 1$ m.

\subsubsection{System Identification}

\begin{figure}[h!]
    \centering
    \includegraphics[width=0.5\columnwidth]{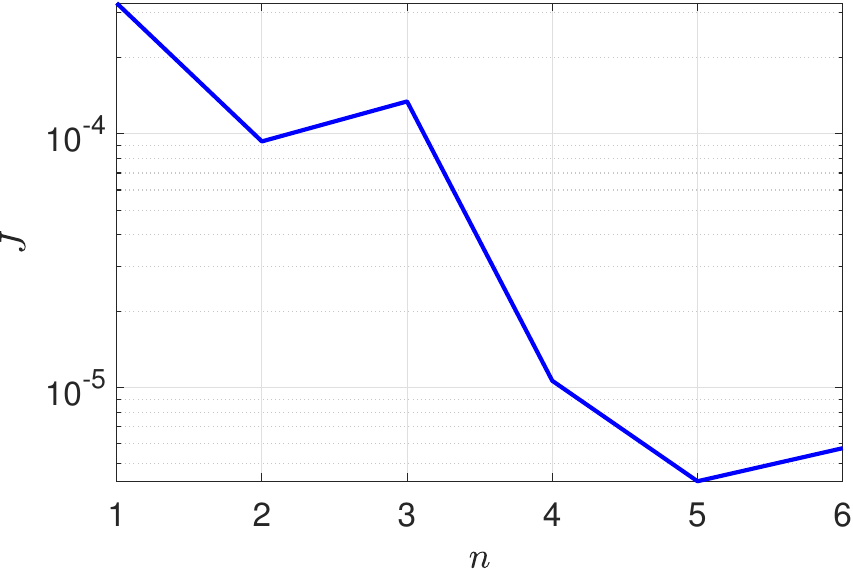}
    \caption{RMSE of identified $G_{yu}$ with order $n$.}
    \label{fig:id_yu}
\end{figure}

A linear model is identified from input-output data, as discussed in Subsections \ref{subsec:sys_id} and \ref{subsec:implementation_details}.
Figure \ref{fig:id_yu} shows the RMSE cost for various choices of $n.$
Although this is not shown in Figure \ref{fig:id_yu}, note that, for $n>6,$ the identified model is unstable and thus $J = \infty.$
Consequently, to design the \hinf controller, we choose the order that minimizes the RMSE, and thus we use the identified 5th-order transfer function, that is, $n = 5.$ 
%

\subsubsection{\hinf Controller Design}
\begin{figure}[h!]
    \centering
    \includegraphics[width=0.5\columnwidth]{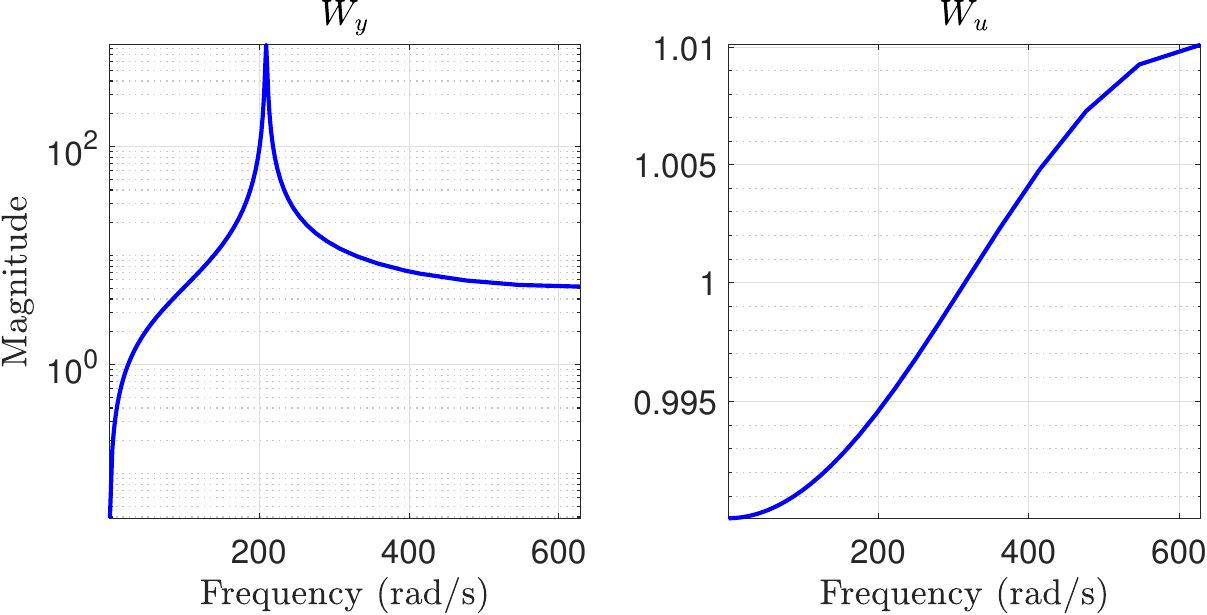}
    \caption{Frequency-weighted filters designed for vibration suppression under external disturbance.}
    \label{fig:bode_dist_filter}
\end{figure}

%
%
%
Aside from the linear model, which was obtained in the previous section, the \hinf controller design procedure discussed in Subsection \ref{subsec:hinf_control_objective} requires the design of filters $W_y$ and $W_u$, which we choose to be
\begin{align}
    W_y(\shiftq) \isdef \frac{7.75\shiftq-7.75}{\shiftq^2-0.99\shiftq+0.98}, \quad W_u(\shiftq) = \frac{1}{\shiftq+0.01}.
\end{align}
%
%
%
%
Figure \ref{fig:bode_dist_filter} shows the frequency response of the filters $W_u(\shiftq)$, and $W_y(\shiftq).$
Note that the filter is designed to magnify the magnitude of the output at the disturbance frequency.
These are used with the system model to obtain a \hinf controller with the MATLAB function \href{https://www.mathworks.com/help/robust/ref/dynamicsystem.hinfsyn.html}.





\subsubsection{Closed-loop Simulations Results}

\begin{figure}[h!]
    \centering
    \includegraphics[width=0.5\columnwidth]{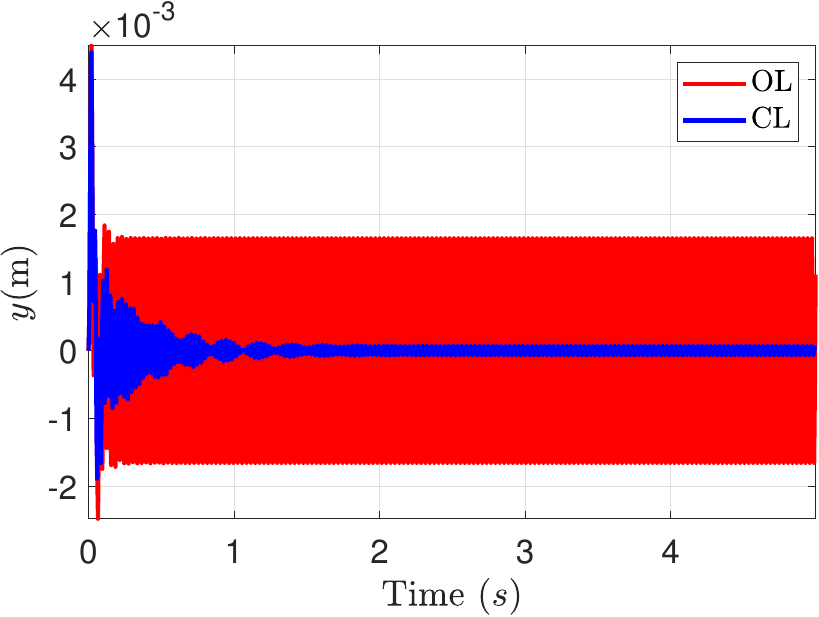}
    \caption{Open-loop (OL) and the closed-loop (CL) responses of the beam under harmonic excitation.}
    \label{fig:closed-loop-hinf}
\end{figure}

%
Figure \ref{fig:closed-loop-hinf} shows the open-loop (OL) and the closed-loop (CL) response of the beam.
Note that, without the control, the amplitude of the tip displacement is approximately $2$ $\rm mm,$
whereas with the controller in the loop, the amplitude of the tip is approximately $0.07$ $\rm mm,$ which is a reduction by a factor of approximately 28. 
%


\subsubsection{Robustness Tests}

\begin{figure}[h!]
    \centering
    \includegraphics[width=1\columnwidth]{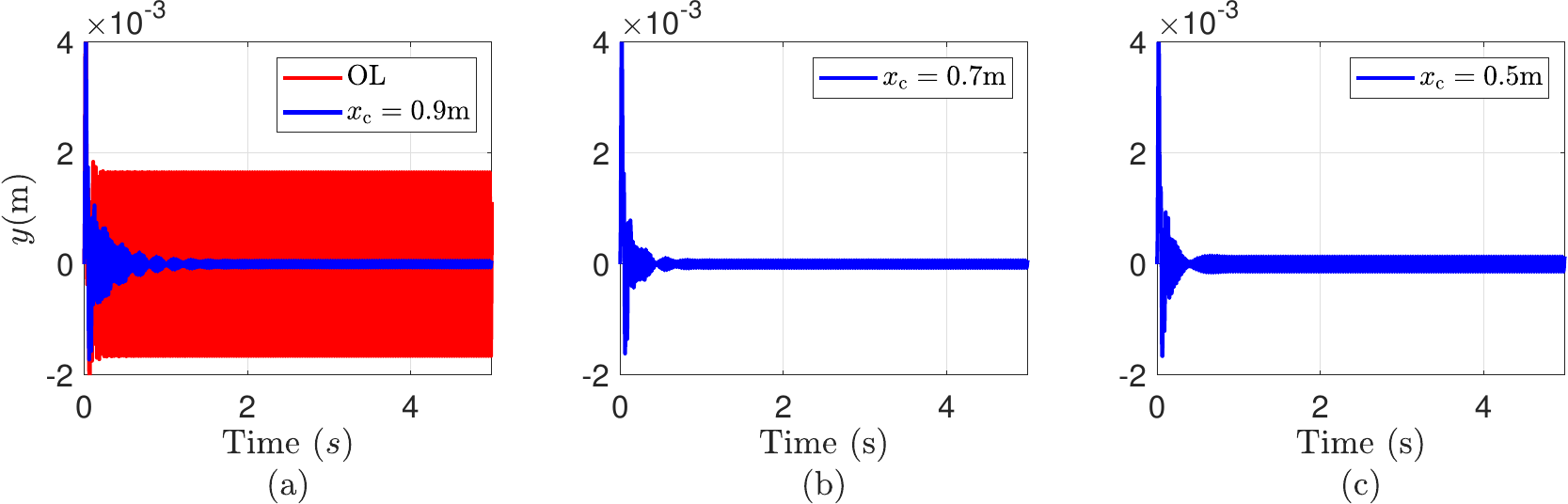}
    \caption{
    Closed-loop response of the beam with various control input locations under harmonic excitation. 
    }
    \label{fig:sensitivy-input-location}
\end{figure}

First, the robustness of the controller is investigated by moving the input location along the beam.
Specifically, we move the input location to $x_\rmc = 0.9 $ m, $x_\rmc = 0.7 $ m, and $x_c = 0.5$ m.
Figure \ref{fig:sensitivy-input-location} shows the closed-loop response in these three scenarios. 
This experiment indicates that the \hinf controller exhibits low sensitivity to changes in actuator placement.

\begin{figure}[h!]
    \centering
    \includegraphics[width=1\columnwidth]{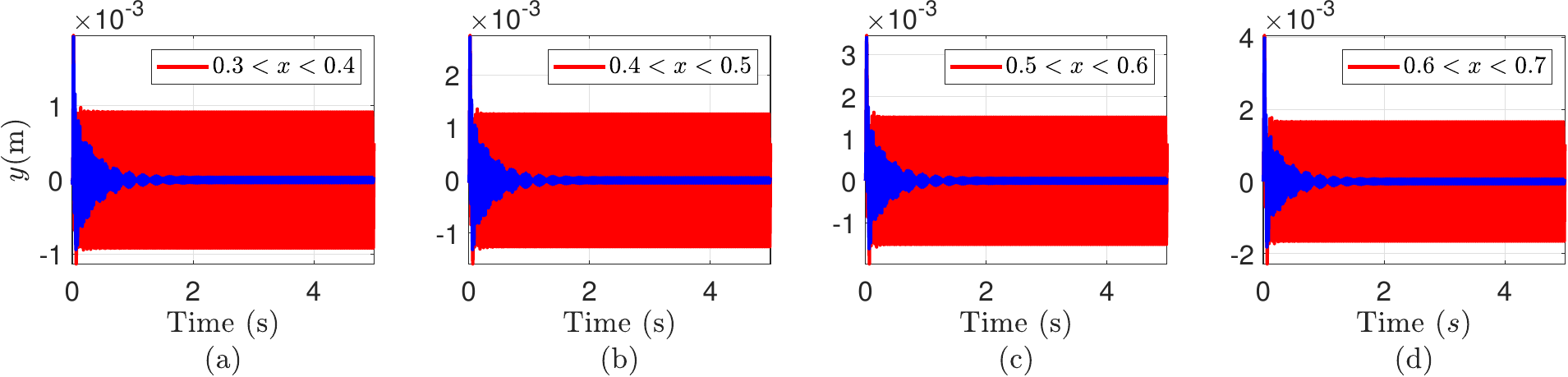}
    \caption{
    Closed-loop response of the beam with various disturbance locations, indicated in the corresponding legends.
    }
    \label{fig:sensitivy-disp-location}
\end{figure}

Next, the robustness of the controller to disturbance location variation is investigated.
For this purpose, $\ell_1$ and $\ell_2$ are varied along the beam.
%
Figure \ref{fig:sensitivy-disp-location} shows the open-loop and closed-loop response with various choices of disturbance locations. 
This experiment indicates that the \hinf controller exhibits low sensitivity to changes in disturbance location.

\subsection{Flutter induced by Aerodynamic Loading}

The aerodynamic loading $p_\rma$ is applied to the cantilever beam to induce self-excited oscillations in the form of flutter.
In the simulations in this subsection, the aeroelastic load with properties described in Table \ref{fig:aeroload_properties} is distributed over the beam, and the sampling time is $T_\rms = 10^{-3}$ s. 
The objective is to reduce the vibrations at the tip of the beam, that is, at $x = L = 1$ m.

\begin{table}[h!]
\centering
\caption{Aeroelastic Load Parameters}
\begin{tabular}{l c c}
\hline
\textbf{Parameter} & \textbf{Symbol} & \textbf{Value} \\
\hline
Free-stream Mach number     & $M_\infty$ & $8$ \\
Flow parameter                 & $\lambda$  & $600$ \\
Mass ratio                     & $\mu$      & $0.1$ \\
Specific heat ratio of air           & $\gamma$   & $1.4$ \\
Free-stream static pressure & $p_\infty$ & $1.88$ \\
\hline
\end{tabular}
\label{fig:aeroload_properties}
\end{table}



\subsubsection{System Identification}

%
%
%
%
%
%

\begin{figure}[h!]
\centering
\includegraphics[width=0.5\columnwidth]{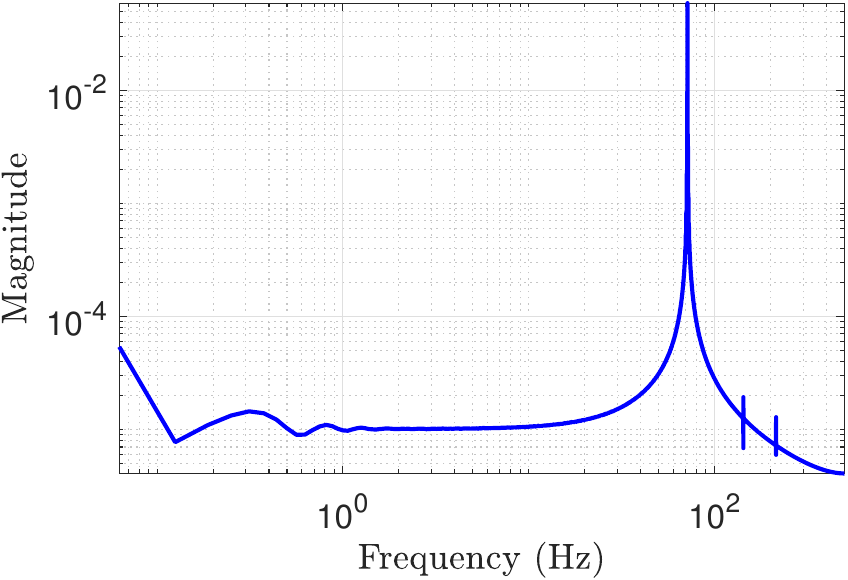}
\caption{FFT of the beam tip displacement under aeroelastic loading and random gaussian excitation at the input.}
\label{fig:fft_aeroload}
\end{figure}

\begin{figure}[h!]
\centering
\includegraphics[width=0.5\columnwidth]{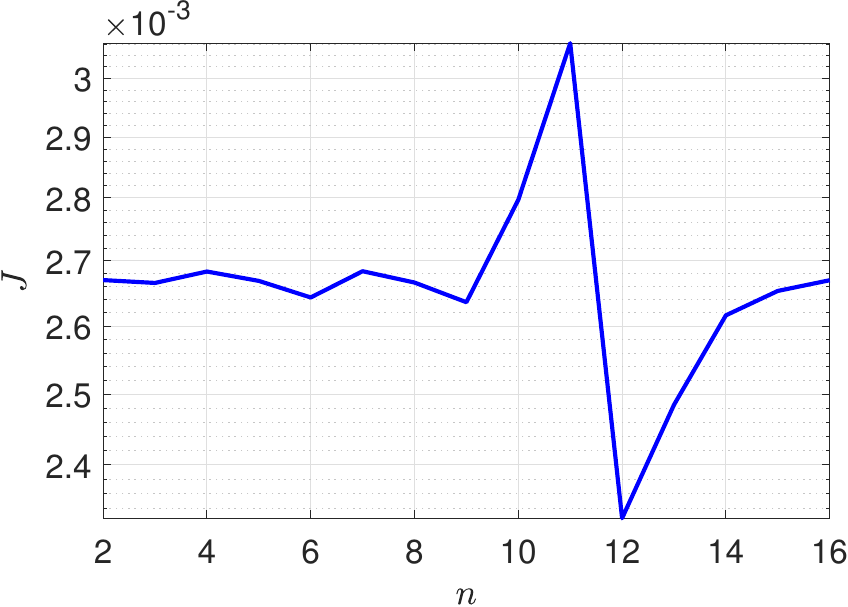}
\caption{RMSE of identified $G_{yu}$ as a function of model order $n$.}
\label{fig:id_yu_aeroload}
\end{figure}

%
%
A linear model is identified from input-output data, as discussed in Subsections \ref{subsec:sys_id} and \ref{subsec:implementation_details}.
In this case, in addition to the random gaussian excitation signal applied to the input, the aerodynamic loading $p_\rma$ is also applied to capture the frequencies corresponding to the aeroelastic oscillations in the input-output data.
The Fast Fourier Transform (FFT) of the output signal is shown in Figure~\ref{fig:fft_aeroload}, which reveals a dominant vibration frequency near $70$ Hz.
Furthermore, Figure~\ref{fig:id_yu_aeroload} shows the RMSE cost for various choices of $n.$
Consequently, to design the \hinf controller, we choose the order that minimizes the RMSE, and thus we use the identified 12th-order transfer function, that is, $n = 12.$ 

\subsubsection{ \hinf Controller Design}

\begin{figure}[h!]
\centering
\includegraphics[width=0.5\columnwidth]{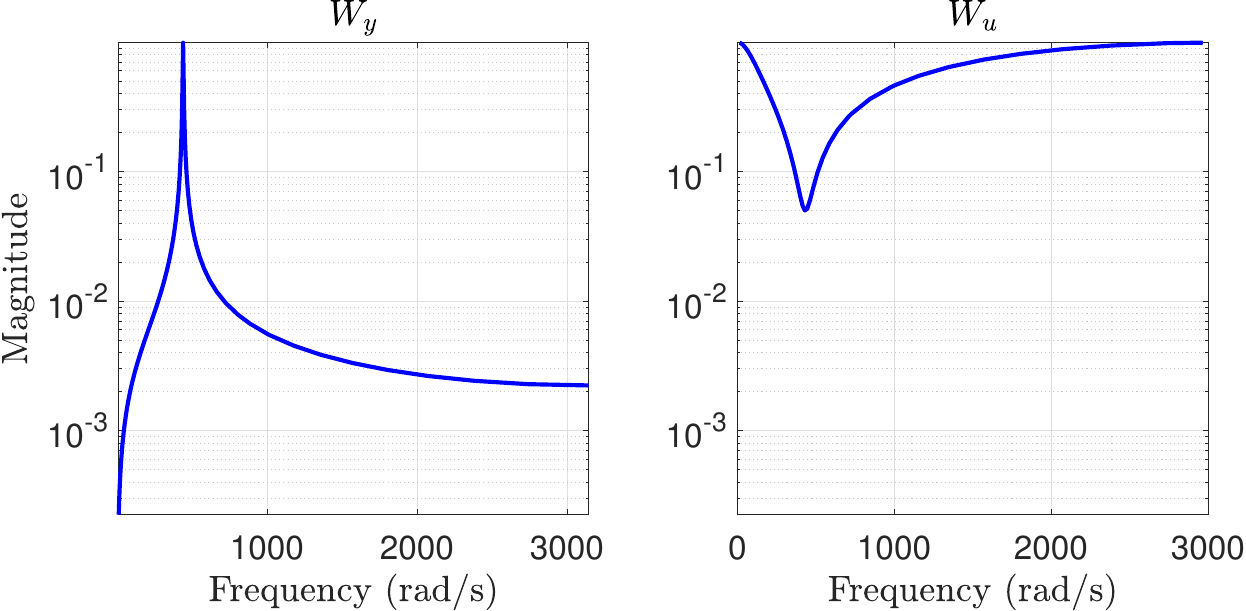}
\caption{Frequency-weighted filters designed for vibration suppression under aeroelastic loading.}
\label{fig:filter_design_aero_load}
\end{figure}

Aside from the linear model, which was obtained in the previous section, the \hinf controller design procedure discussed in Subsection \ref{subsec:hinf_control_objective} requires the design of filters $W_y$ and $W_u$.
Motivated by the dominant frequency observed in Figure \ref{fig:fft_aeroload}, the frequency-weighted input and output filters are designed to attenuate vibrations around this dominant frequency.
The resulting filters are
\begin{align}
W_y(\mathbf{q}) &= \frac{0.0042\mathbf{q} - 0.0042}{\mathbf{q}^2 - 1.806\mathbf{q} + 0.995},
\qquad
W_u(\mathbf{q}) = \frac{0.566\mathbf{q}^2 - 0.987\mathbf{q} + 0.52}{\mathbf{q}^2 - 0.987\mathbf{q} + 0.087}.
\end{align}
%
%
Figure ~\ref{fig:filter_design_aero_load} shows the frequency response of the filters $W_y (\shiftq)$ and $W_u (\shiftq).$
As seen in Figure ~\ref{fig:filter_design_aero_load}, the design increases the weighting on the output near the dominant vibration frequency, while allowing the control input to become more aggressive in that same frequency range.
This ensures that the controller effectively suppresses the aeroelastic mode without excessively penalizing the required control effort.

\subsubsection{Closed-loop Simulations}

\begin{figure}[h!]
\centering
\includegraphics[width=0.5\columnwidth]{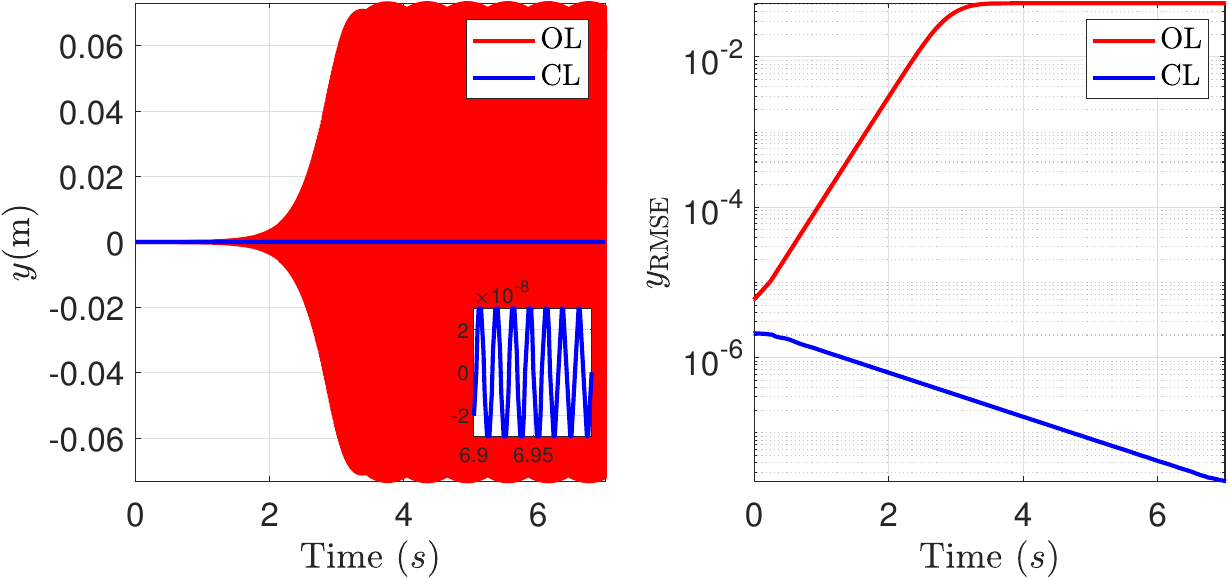}
\caption{Open-loop (OL) and closed-loop (CL) response of the beam under aeroelastic loading.}
\label{fig:bode_dist_filter_aero_load}
\end{figure}

Figure~\ref{fig:bode_dist_filter_aero_load} shows both the open-loop (OL) and closed-loop (OL) responses of the cantilever beam under aeroelastic loading.
The closed-loop response corresponds to the system controlled by the designed \hinf controller.
In both cases, the system is excited at $t = 0$ s by an impulse to create the initial oscillations.

As shown in Figure~\ref{fig:bode_dist_filter_aero_load}, the initial perturbation causes the system to oscillate significantly in the open-loop case.
However, in the case where the controller is applied, the vibrations induced by the aeroelastic load are effectively suppressed.

\subsubsection{Robustness Tests}

\begin{figure}[h!]
\centering
\includegraphics[width=\columnwidth]{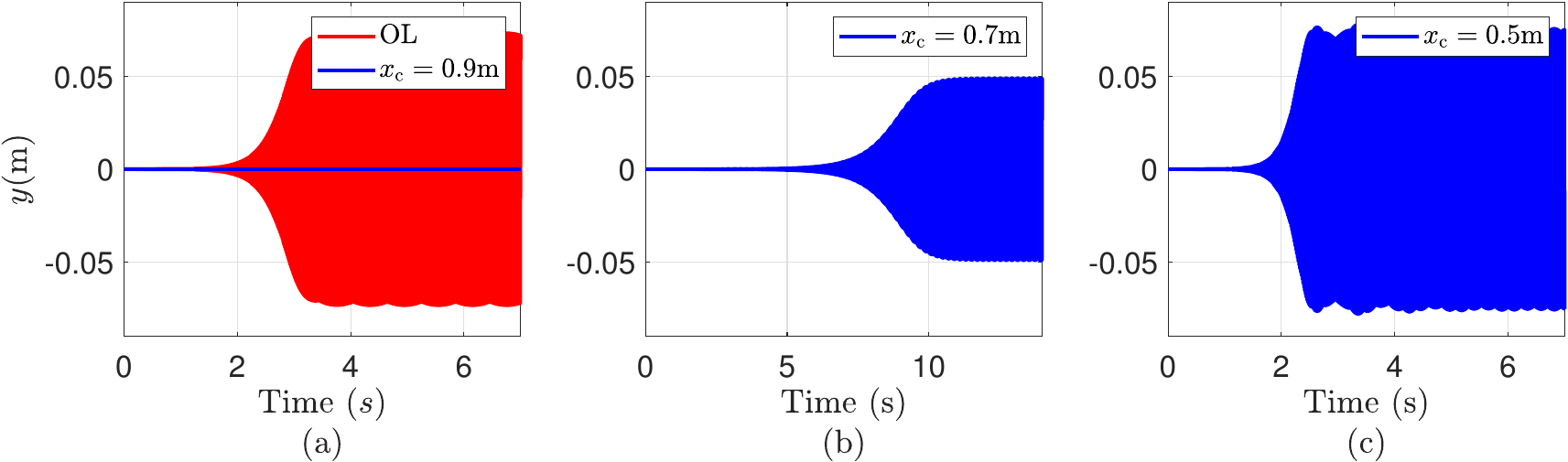}
\caption{Closed-loop response of the beam under aeroelastic loading with input applied at different locations.}
\label{fig:sensitivy-input-location_aeroload}
\end{figure}

Next, we vary the location of the applied input along the beam to evaluate the robustness of the designed controller with respect to changes in actuation position.
In the previous test, the input was applied at the tip of the beam.
In Figure~\ref{fig:sensitivy-input-location_aeroload}, we shift the input location to $x_\rmc=0.9$ m, $x_\rmc=0.7$ m, and $x_\rmc=0.5$ m along the beam and examine the resulting closed-loop responses. 
As the results show, the controller continues to provide displacement attenuation at the tip of the beam for input locations up to $x_c = 0.7$ m, after which its performance begins to degrade.

\section{Conclusions and Future Work}
\label{sec:conclusions}

This paper presented the development and validation of an \hinf output-feedback controller for aeroelastic vibration suppression in a cantilevered beam.
A finite element model (FEM) is first developed to simulate the nonlinear structural dynamics. 
Then, a low-order linear system was identified from random gaussian input response data to synthesize the \hinf controller. 
Frequency-weighted dynamic filters were used to target suppression at a dominant disturbance frequency.
A preliminary study was performed by testing the stabilization performance of the \hinf controller in the cantilevered beam FEM model under harmonic excitation.
In this case, simulation results demonstrate that the proposed controller significantly reduces tip displacement amplitude by a factor of approximately 28 under the controller design conditions.
Furthermore, robustness tests show that the controller maintains its performance across various actuator and disturbance locations, indicating low sensitivity to spatial placement. 
%
%
Next, an \hinf controller was employed to suppress oscillations induced by aerodynamic loading. The redesigned \hinf controller effectively attenuates these oscillations.
However, its performance was found to be sensitive to actuator placement. 
Future work will focus on developing learning-based control techniques for aeroelastic vibration suppression that are robust to variations in sensor and actuator locations.

%
%
%


\bibliography{Bib/structure}
\end{document}

%% file: PackagesCommands.tex
\usepackage{amsmath} 
\usepackage{amsfonts}

\usepackage{float} 

\usepackage{subcaption}

\usepackage{tikz}
\usetikzlibrary{circuits.ee.IEC}
\usetikzlibrary{shapes,arrows,calc,positioning}
\tikzstyle{bigblock} = [draw, fill=blue!20, rectangle, 
    minimum height=6em, minimum width=8em]
\tikzstyle{medblock} = [draw, fill=blue!20, rectangle, 
    minimum height=4em, minimum width=4em]    
\tikzstyle{mux} = [draw, fill=black!20, rectangle, 
    minimum height=5em, minimum width=0.1em]    
\tikzstyle{smallblock} = [draw, fill=blue!20, rectangle, 
    minimum height=2em, minimum width=3em]
    
\tikzstyle{data_block} = [draw, fill=green!20, rectangle, 
    minimum height=2em, minimum width=3em]
\tikzstyle{ops_block} = [draw, fill=blue!20, rectangle, 
    minimum height=2em, minimum width=3em]    
\tikzstyle{est_block} = [draw, fill=red!20, rectangle, 
    minimum height=2em, minimum width=3em]    
    
\tikzstyle{sum} = [draw, fill=blue!20, circle, node distance=1cm,minimum height=0.5cm]
\tikzstyle{signal} = [coordinate]
\tikzstyle{pinstyle} = [pin edge={to-,thin,black}]
\tikzstyle{block} = [draw, fill=blue!20, rectangle, 
    minimum height=3em, minimum width=9em]
\tikzstyle{blockS} = [draw, fill=blue!20, rectangle, 
    minimum height=3em, minimum width=4em]    
\tikzstyle{input} = [coordinate]
\tikzstyle{output} = [coordinate]

\tikzstyle{gain} = [draw, fill=blue!20, regular polygon, regular polygon sides=3, shape border rotate=30]
\tikzstyle{gain_vert} = [draw, fill=blue!20, regular polygon, regular polygon sides=3, shape border rotate=0]

\usetikzlibrary{matrix}

\usetikzlibrary{positioning}
\usetikzlibrary{math}

\usepackage{hyperref}
\usepackage{xcolor}
\hypersetup{
    colorlinks,
    linkcolor={blue!100!black},
    citecolor={blue!50!black},
    urlcolor={blue!80!black}
}

\newcommand{\bc}{\begin{center}}
\newcommand{\ec}{\end{center}}
\newcommand{\benum}{\begin{enumerate}}
\newcommand{\eenum}{\end{enumerate}}

\newcommand{\matl}{\left[ \begin{array}}
\newcommand{\matr}{\end{array} \right]}

\renewcommand{\matl}{\begin{bmatrix}}
\renewcommand{\matr}{\end{bmatrix}}

\newcommand{\matls}{\left[ \begin{smallmatrix}}
\newcommand{\matrs}{\end{smallmatrix} \right]}
\newcommand{\isdef}{\stackrel{\triangle}{=}}

\newcommand{\inv}{^{-1}}

\newcommand{\rmF}{{\rm F}}

\newcommand{\rmL}{{\rm L}}

\newcommand{\rmN}{{\rm N}}

\newcommand{\rmT}{{\rm T}}

\newcommand{\rma}{{\rm a}}
\newcommand{\rmb}{{\rm b}}
\newcommand{\rmc}{{\rm c}}
\newcommand{\rmd}{{\rm d}}
\newcommand{\rme}{{\rm e}}
\newcommand{\rmf}{{\rm f}}

\newcommand{\rms}{{\rm s}}

\newcommand{\BBR}{{\mathbb R}}

\newcommand{\SG}{{\mathcal G}}
\newcommand{\SH}{{\mathcal H}}

\newcommand{\shiftq}{{\textbf{\textrm{q}}}}

\usepackage{enumitem,amssymb}
\newlist{todolist}{itemize}{2}
\setlist[todolist]{label=$\square$}
\usepackage{pifont}